\documentclass[aps,twocolumn,showpacs]{revtex4}
\usepackage{graphicx}
\begin{document}
\title{Delocalization transition in an exactly solvable many-body system in Two Dimensions}
\author{Ganesh R}
\affiliation{Birla Institute of Technology and Science, Pilani,
India.}
\author{Saugata Ghosh}\email{sghosh@ictp.it}
\affiliation{The Abdus Salam ICTP, Strada Costiera 11, 34100,
Trieste, Italy.}
\date{\today}
\begin{abstract}
We study delocalization transition in a many body system in two
dimension. We identify the presence of a complex vector potential
that gives rise to this transition.
\end{abstract}
 \maketitle{}

Non-Hermitian Hamiltonians  appear in the study of quantum dots
coupled to leads where it is convenient to exclude the degrees of
freedom in the leads by integrating over them. This scheme results
in an effective non-Hermitian Hamiltonian whose eigenstates
contain an imaginary part that accounts for the decay into the
leads.

In Ref. \cite{hatano}, Hatano and Nelson, considered a different
kind of non-Hermitian Hamiltonians, one with a constant imaginary
vector potential. This leads to the appearance of a first order
derivative in the Hamiltonian that, as pointed out by Efetov
\cite{efetov} , can be assimilated to the appearance of a certain
preferential direction in space. The introduction of this
ingredient gives rise to new effects such as a localization
transition in a disordered one dimensional system
\cite{beenakker}.

In this context, we study analytically an exactly solvable
$N$-body system in two-dimensions, which shows a localization to
delocalization transition even in the absence of random potential.
The hamiltonian is non-hermitian with a complex vector potential.

We derive the ground state eigenvalue and eigenfunction for such
systems and extract several interesting properties by identifying
the square of the radial wave function with the joint probability
distribution (JPD) of eigenvalues of the Laguerre ensembles of
random matrix theory (RMT). We also study the particle density and
pair correlation function (PCF) for such systems. We observe
``universality'' in the PCF as expected for such systems \cite{gp}
from random matrix results.

Let us consider a two-dimensional system of $N$ particles in the
$x-y$ plane, subject to a vector potential $\vec
A=[x\hat{x}+y\hat{y}]/[2(x^2+y^2)]$. The single body Hamiltonian
is given by
\begin{equation}
H_{sp}={\left[(\hbar/i)\nabla+i\vec {A}\right]}^2.
\end{equation}
The many body Hamiltonian is given by
\begin{eqnarray}
H=\sum_{j}\{[(\hbar/i)\nabla+i\vec {A_j]}^2 ,
\end{eqnarray}
where $A_j$ is the vector potential.

Making the transformation  $x=r\cos\theta$ and $y=r\sin\theta$ and
taking $\hbar=m=1$, we get
\begin{eqnarray}\label{phamiltonian}
 H=-\sum_{i}\frac{{\partial}^2}{{\partial}r_{i}^2}
    -\sum_{i}\frac{1}{r_{i}^2}
    (\frac{{\partial}^2}{{\partial}\theta_{i}^2}+
    \frac{1}{4}).
    \end{eqnarray}

Here, we have used the relation
\begin{equation}
\nabla^2=\frac{\partial^2}{\partial
r^2}+\frac{1}{r^2}\frac{\partial^2}{\partial\theta^2}+\frac{1}{r}\frac{\partial}{\partial
r}.
\end{equation}
In terms of the polar coordinate, the vector potential can be
written as
\begin{eqnarray}
\nonumber \vec A &=&
-\frac{1}{2}{\nabla}_{\theta}\hat{\theta},\\
&=& \frac{\hat{r}}{2r}
\end{eqnarray}
where,
\begin{eqnarray}
{\nabla}_{\theta}=\frac{1}{r}\frac{\partial}{\partial
\theta},\hspace{1cm}{\nabla}_{r}=\frac{\partial}{\partial r},
\end{eqnarray}
and
\begin{eqnarray}
\hat{r}=\hat{x}\cos\theta+\hat{y}\sin\theta ,\hspace{1cm}
\hat{\theta}=-\hat{x}\cos\theta+\hat{y}\sin\theta,
\end{eqnarray}
where $\hat{x}$, $\hat{y}$, $\hat{r}$ and $\hat{\theta}$  are unit
vectors along the $x$, $y$, $r$ and $\theta$ directions
respectively. This expression in turn allows us to view the
complex vector potential as some velocity operator acting along
the angular coordinate. This maybe the reason for which one can do
away with the random part of the potential (as is seen in the work
of previous authors), but retaining the effect of disorder.
However, one needs to understand this aspect more clearly.

Replacing this in the Schrodinger equation
\begin{equation}
H {\Psi}_n(r_i,{\theta}_i)=E_n{\Psi}_n(r_i,{\theta}_i),
\end{equation}
we obtain the ground state eigenfunction and eigenvalue.

Let us take the solution wavefunction of the form
$\Psi_{n}(r_{i},\theta_{i})=R(r_{1},\ldots,r_N)\Theta(\theta_{1},\ldots,\theta_N)$.
Using the separation of variable and equating the angular part to
zero, we get

\begin{eqnarray}\label{radial}
-\sum_{i}\frac{{\partial}^2}{{\partial}r_{i}^2}R(r_{1},\ldots,r_{N})
=E_{n}R(r_{1},\ldots,r_{N})
\end{eqnarray}

The angular part of the equation is given by
\begin{equation}
\frac{{\partial}^2}{\partial{\theta_{i}}^2}\Theta(\theta_{1},\ldots,\theta_{N})=-\frac{1}{4}\Theta
\end{equation}

Now, let us assume that the solution wavefunction has the form
\begin{eqnarray}
\label{radialsol}
R_{n}(r_{i})=\prod_{i<j}{(r_{i}-r_{j})^\lambda}\exp(-{\alpha}_{k}\sum_{i}r_{i})\\
\nonumber
\Theta(\theta_{1},\ldots,\theta_{N})=A\exp(\frac{\pm i}{2}\sum_{i}\theta_{i})\\
\end{eqnarray}
where $\lambda$ can take only two values, i.e. $0$ and $1$. Then
$\lambda=0$ will correspond to bosons while $\lambda=1$ to
fermions. In that case, $\alpha_{k}=\sum[\kappa \mp ik_{n}]$ and
the eigenvalue for the $N$-body Hamiltonian will be given by
\begin{equation}
\label{eigen} E_{k}=-N({\kappa}^2-{k_{n}}^{2}+2ik_{n}\kappa),
\end{equation}
We must note that for a given system $\kappa$ is constant. The
ground state (the state that minimizes the real part of $E_{n})$
is localized for $k_{n}=0$. The localized ground state is
$E_0=-N{\kappa}^2$. In the extended state, in order to have a well
defined wave function, we need to specify the boundary conditions
specific to the system. The quantity $\kappa$ is a measure of the
amount of localization.

Now, we will prove that (\ref{radialsol}) is indeed a solution of
the radial equation (\ref{radial}). Let us take
\begin{eqnarray}
\phi=\prod_{i<j}(r_{i}-r_{j})^\lambda\\
\varphi=\exp(-{{\alpha}_{k}}\sum_{i}r_{i})
\end{eqnarray}
It can be easily seen that
\begin{eqnarray}
\sum_{i}\frac{{\partial}^2}{{\partial}r_{i}^2}(\phi\varphi)=
\sum_{i}\left[\phi\frac{{\partial}^2\varphi}{{\partial}r_{i}^2}
+2\frac{{\partial}\phi}{{\partial}r_{i}}\frac{{\partial}\varphi}{{\partial}r_{i}}
+\varphi\frac{{\partial}^2\phi}{{\partial}r_{i}^2}\right]\\
\end{eqnarray}
Substituting Eqs.(14,15), the first term gives
\begin{equation}
\phi\frac{{\partial}^2\varphi}{{\partial}r_{i}^2}=
N({\kappa}^{2}-{k_{n}}^{2}+2ik_{n}\kappa)
\end{equation}
while
\begin{equation}
\varphi\frac{{\partial}^2\phi}{{\partial}r_{i}^2}=
2({\lambda}^{2}-\lambda)\sum_{i<j}\frac{1}{{(r_{i}-r_{j})}^{2}}\phi\varphi.
\end{equation}
The middle term becomes zero due to symmetry. We can also see that
for Eq.[\ref{radialsol}] to be a solution of the radial part of
the Hamiltonian, $\lambda$ can only be $0$ or $1$. Thus we have
proved Eq.[\ref{radialsol}]. The proof of the angular part is
straightforward.

It is interesting to note from Eq.(\ref{eigen}) that for
$k_{n}=0$, the wavefunction is exponentially localized and the
eigenvalue does not contain any imaginary part. However, as
$k_{n}$ increases, the wavefunction becomes extended in nature and
a complex part is added to the eigenvalue.

It is at this point that, we will write $\psi_{n}$  as
\begin{equation}
\psi_{n}=C^{1/2}\prod_{i<j}{|r_{i}-r_{j}|}^{\lambda}
\exp[\pm\alpha_{k}r_{i}]\exp[\pm\frac{i}{2\theta}],
\end{equation}
Defining the scalar product $(\psi_{n},\psi_{m})$ as
\begin{eqnarray}
\nonumber
(\psi_{n},\psi_{m}) &=& C\int\int\prod {|r_{i}-r_{j}|}^{\beta}\times\\
\nonumber
&& \exp(-\kappa\sum (r_i)\pm i(k_n-k_m)\sum r_i)\\
&=&{\pi}^{N}C\delta_{nm}.
\end{eqnarray}
Here for $n\neq m$, the term on the right is highly oscillatory
and hence goes to zero. For $n=m$, we get the normalization. Here,
$C$ is the well known Selsberg constant  given by
\begin{equation}
C^{-1}=\prod_{j=0}^{N-1}\frac{\Gamma(1+\lambda+j\lambda)\Gamma(1+j\lambda)}{\Gamma(1+\lambda)}.
\end{equation}
Then ${|{\psi}_{n}|}^{2}$ is given by
\begin{equation}
{|{\psi}_{n}|}^{2}=C\prod_{i<j}{|r_{i}-r_{j}|}^{\beta}\exp(-2\kappa\sum(r_i),
\end{equation}
with $\beta=2\lambda $, and $C$ being the normalization constant.

Now one may  interpret $|{\psi_{n}}|^{2}$ to be identical with the
j.p.d. of Laguerre ensembles of random matrices \cite{gp}. We
define the $m$-particle correlation function

\begin{equation}
Y_{m}(r_{1}, r_{2},\ldots, r_{m})=\int\ldots\int dr_{m+1}\ldots
dr_{N}d{\theta}_{m+1}\ldots d{\theta}_{N}{|\psi_{0}|}^{2}.
\end{equation}
as the probability of finding $m$-particles in the intervals
$r_{i}$ and $r_{i}+\Delta r_{i}$ and $\theta_{i}+\Delta
\theta_{i}$, irrespective of the position of the other particles.
$m=1$ and $2$ correspond to  the particle density and PCF
respectively.

It has also been shown in Ref. \cite{gp} that $Y_{1}(x)$
corresponds to the density of zeros of the polynomial having
weight function given by that of the Laguerre Polynomial. It has
been shown by Dyson in the context of random matrices that for
$\beta=1$, $2$ and $4$,
 $Y_{m}$ can be written in terms of orthogonal
and skew-orthogonal polynomials. For $\beta=2$, corresponding to
the Fermionic solution, the PCF can be written as
\begin{equation}
\label{Y2}
Y_{2}(r_{1},r_{2})=\sum_{\mu=0}^{N-1}{h_{\mu}}^{-1}[q_{\mu}(r_{1})q_{\mu}(r_{2})]
\exp[-2\kappa r_{1}]
\end{equation}
where $q_{\mu}(r)$ are orthogonal polynomials corresponding to the
normalization condition
\begin{equation}
\int_{-\infty}^{\infty}q_{\mu}(y)q_{\nu}(y) \exp[-2\kappa
r]dy=h_{\mu}\delta_{\mu\nu}.
\end{equation}

It should be noted that our choice of circular coordinate has
considerably simplified our calculation as $|{\psi}|^{2}$ does not
contain $\theta$ explicitly. Using the Christoffel-Darboux formula
in Eq.[\ref{Y2}], we can perform the sum. Finally using the
asymptotic form of the Laguerre polynomials, and taking the limit,
we get for the particle density. Here, we will give an alternate
derivation (used previously in \cite{gp}) for the derivation of
particle density.

We define the resolvent
\begin{equation}
G(z)=\int\frac{Y_{1}(r)}{z-r}dr,
\end{equation}
which satisfies
\begin{equation}
G(x+i0)=\int\frac{Y_{1}(r)}{x-r}dr-i\pi Y_{1}(r).
\end{equation}
Also, from the definition of ${|\psi_{n}|}^{2}$, we may write
\cite{gp}
\begin{eqnarray}
\frac{\partial Y_{1}(r)}{\partial r}&=&
\beta\int\frac{Y_{2}(r,r_{2})}{(r-r_{2})}dr_{2}-2\kappa Y_{1}(r).
\end{eqnarray}
For large $N$, replacing $Y_{2}(r,r_{2})\simeq Y_{1}(r)Y_{1}(r_2)$
and dropping $\partial Y_{1}(r)/\partial r$, we get
\begin{equation}
-2\kappa\int dr\frac{rY_{1}(r)}{(z-r)} +\frac{\beta}{2}\int \int
drdr_{2}\frac{zY_{1}(r)Y_{1}(r_{2})}{(z-r)(z-r_{2})} = 0.
\end{equation}
This gives
\begin{equation}
\beta zG^{2}(z)+4\kappa N-4\kappa ZG(z)=0,
\end{equation}
solving which we get for the particle density
\begin{eqnarray}
\label{density} \pi\beta Y_{1}(r) &=&
2\kappa\sqrt{\frac{N\beta/\kappa-r}{r}}.
\end{eqnarray}

 It is interesting to note that as pointed out earlier,
$\kappa$ is a measure of the degree of localization. For $\kappa$
large compared to $N$, particles get more strongly localized
around the origin. However, as $\kappa$ decreases, particles start
getting delocalized, until they form extended states. Also, it
should be noted that in the thermodynamic limit, the particle
density, after proper scaling, can be made independent of $\beta$
and hence applicable for both boson and fermion. However, in this
case, the nature of localization is exponential for bosons
($\lambda=0$), while it is given by Eq.[\ref{density}] for
fermions ($\lambda=1$).

To calculate  the PCF, essential for the study of thermodynamic
properties,  for $\beta=2$, it has been proved in Ref.\cite{gp}
that for interparticle spacing $(r_{1}-r_{2})\equiv \Delta
r\rightarrow 0$, $N\rightarrow \infty$ and defining $r'=\Delta
r.Y_{1}(r)$, the scaled PCF is universal. For $\beta=2$, it is
given by
\begin{equation}
\label{corr} Y_{2}(r_{1},r_{2})=1-\frac{\sin^2(\pi r')}{(\pi
r')^2}.
\end{equation}
This shows that the probability of finding two particles in the
same state ($\Delta r=0$) is zero. This is in agreement with the
Pauli exclusion principal, applicable for a system of fermions.

Thus we have studied the localization to delocalization transition
 in a system of $N$-particles in the presence of a
complex vector potential. For localized state, particles tend to
gather around the origin, as shown in Eq.[\ref{density}]. At the
same time, the PCF shows that the particles are not allowed to
condense to one position. The corresponding eigenvalue is real.
However, for delocalized state, we observe that the eigenvalue has
complex part.

S.G. is  grateful to Dr. Louis Foe Torres for many useful
discussions.

\end{document}